\documentclass{article}

\usepackage{spconf,amsmath,graphicx,hyperref}
\usepackage{amsfonts}
\usepackage{booktabs}
\usepackage{svg}
\ninept
\title{Neural personal sound zones with flexible bright zone control}
\name{Wenye Zhu ${ }^{1,2}$, Jun Tang${ }^{2}$, Xiaofei Li ${ }^{2,*}$\thanks{* corresponding author}}
\address{${}^1$ Zhejiang University, Hangzhou, China\\
	${}^2$ Westlake University \& Westlake Institute for Advanced Study, Hangzhou, China}
\begin{document}
%\ninept
%
\maketitle
\begin{abstract}
Personal sound zone (PSZ) reproduction system, which attempts to create distinct virtual acoustic scenes for different listeners at their respective positions within the same spatial area using one loudspeaker array, is a fundamental technology in the application of virtual reality. For practical applications, the reconstruction targets must be measured on the same fixed receiver array used to record the local room impulse responses (RIRs) from the loudspeaker array to the control points in each PSZ, which makes the system inconvenient and costly for real-world use. In this paper, a 3D convolutional neural network (CNN) designed for PSZ reproduction with flexible control microphone grid and alternative reproduction target is presented, utilizing the virtual target scene as inputs and the PSZ pre-filters as output. Experimental results of the proposed method are compared with the traditional method, demonstrating that the proposed method is able to handle varied reproduction targets on flexible control point grid using only one training session. Furthermore, the proposed method also demonstrates the capability to learn global spatial information from sparse sampling points distributed in PSZs.
% reducing the cost of recording, describing and storing of these variable virtual scenes to be reproduced by the system.

\end{abstract}
\section{Introduction}
\label{sec:intro}
% Definition/background
Personal sound zones refer to individualized acoustic spaces in which listeners experience a predesigned or personally preferred acoustic scene \cite{druyvesteyn1997personal}. A PSZ system aims to provide multiple listeners with their own PSZs within the same acoustic environment, employing a loudspeaker array with well-designed pre-filters to render the audio signals. Two types of spatial sound zones are usually considered: the bright zone (BZ) and the dark zone (DZ). The BZ denotes the area where the target signal is reproduced, whereas the DZ denotes the area where the signal is suppressed or constrained to a low level. Various applications of PSZ systems have been studied and realized, including personal computers and televisions \cite{chang2009realization}, car cabins \cite{cheer2013design, Vindrola2021adapt},  mobile devices \cite{elliot2010minimally}, domestic environments \cite{jacobsen2023domestic}and hospital setting \cite{fangel2023hospital}. 

% multipoint-optimization-based method: 
% conception： control points， ATF
Commonly used PSZ techniques include acoustic contrast control (ACC) \cite{choi2002illuminate} and pressure matching (PM) \cite{druyvesteyn1997personal}. In both methods, the sound zones are discretized into a set of points, referred to as control points, where the sound pressures or acoustic transfer functions (ATFs) are controlled by the pre-filters acting on the loudspeaker array. PM minimizes the reproduction errors of ATFs at control points through a regularized least squares method, while ACC maximizes the contrast between the sound energy in the BZ and the DZ. Further studies expand these methods from frequency domain to time domain \cite{elliot2012robustness, cai2014tdacc, simon2015tdpm} and subband domain \cite{so2019subband, moles2020subband, tang2025relax}. In addition, a generalized hybrid framework termed variable span trade-off (VAST) was introduced in \cite{VAST1} and further modified in \cite{VAST2, VAST3, VAST4}, with ACC and PM identified as two special cases. 
More recently, some deep learning–based techniques have offered alternatives for PSZ. In \cite{PePe2020DNNPSZ, Pepe2022DNNPSZ}, two non–data-driven neural networks for PSZ were developed, aimed at producing a flat frequency response in the BZ and an all-zero response in the DZ through a neural network combined with a tailored loss function. 
In \cite{YQ2025DNNPSZ}, a spatially adaptive neural network was trained for head-tracked PSZ rendering, achieving robustness comparable or superior to traditional methods while offering better efficiency for real-time use. 
% Moreover, a convolutional neural network (CNN) for conventional sound reproduction task (a simplified PSZ problem by omitting the requirement of suppressing energy in the dark zone) was developed, which provided valuable insights for network architecture design in PSZ \cite{hong2023reproduction}.
% The methods mentioned above all require a fixed control point grid for control target measurement as the PSZ system input, meaning that in practical applications, such as AR systems with immersive audio, the microphone array recording remote acoustic scenes must maintain the same topology and number of elements as the array used to measure the local RIRs from the loudspeakers to the control points. This constraint makes the practical deployment of PSZ systems inconvenient.

For practical applications such as AR systems with immersive audio, where the goal is to reproduce a specific real-world acoustic scene in the BZ, the aforementioned methods typically require a fixed control point grid to measure both the target ATF in the remote scene and the local RIRs (from the loudspeaker array to the control points) to prevent mismatches. In other words, whenever the target scene changes, the microphone array used to capture it must maintain the same grid pattern as in the original RIR measurement in the local room. This constraint makes the reproduction of varying acoustic scenes costly and inconvenient. Considering the persistent learnable characteristics of deep learning techniques, we think the neural network holds the potential to address this challenge, and some attempts has been made in the related area. \cite{lluis2020inpaint} proposed a deep-learning-based method for sound field reconstruction, capable of performing inpainting and super-resolution using very few microphones with irregular distributions.
\cite{hong2023reproduction} proposed an end-to-end sound field reproduction model using sparse convolutional layers, which can generate loudspeaker driving signals from microphone sound-pressure signals to reproduce target scene in a specific area. 
% In \cite{YQ2025DNNPSZ}, a spatially adaptive neural network was trained for head-tracked PSZ rendering, achieving robustness comparable or superior to traditional methods while offering better efficiency for real-time applications. 
Motivated by the limitation of fixed grid framework and inspired by the work of \cite{lluis2020inpaint} and \cite{hong2023reproduction}, this work develops an end-to-end Neural PSZ model which supports flexible control grid pattern while requiring fewer control points than the PM method. The proposed model uses 3D CNN framework to learn the spatial information within the sound zone. The inputs of the model are the acoustic transfer function (ATF) of the target sound scenes in each zone, and the outputs are the set of pre-filters for the loudspeaker array. The loss of network is the difference between the ATF of the ground truth target and the ATF reproduced in PSZs. 
Although the ultimate goal is to apply the model to reprudction of PSZs with arbitrary remote acoustic scene and grid patterns, the designing and training of a complex system with multiple degrees of flexibilty is extremely challenging. To simplify the problem, this work only focus on validating the feasibility of a CNN to support alternative target, variable control grid patterns and sparser control grids, and our experiments consider a simplified scenario, i.e., reproducing an arbitrary virtual source in a local room. 

In the following sections, we first formulate the PSZ problem, then introduce the proposed Neural PSZ system and CNN model. The experimental results are also shown and discussed.

\section{Problem formulation}
\label{sec:problem}

\begin{figure}
    \centering
    \includegraphics[width=0.9\linewidth]{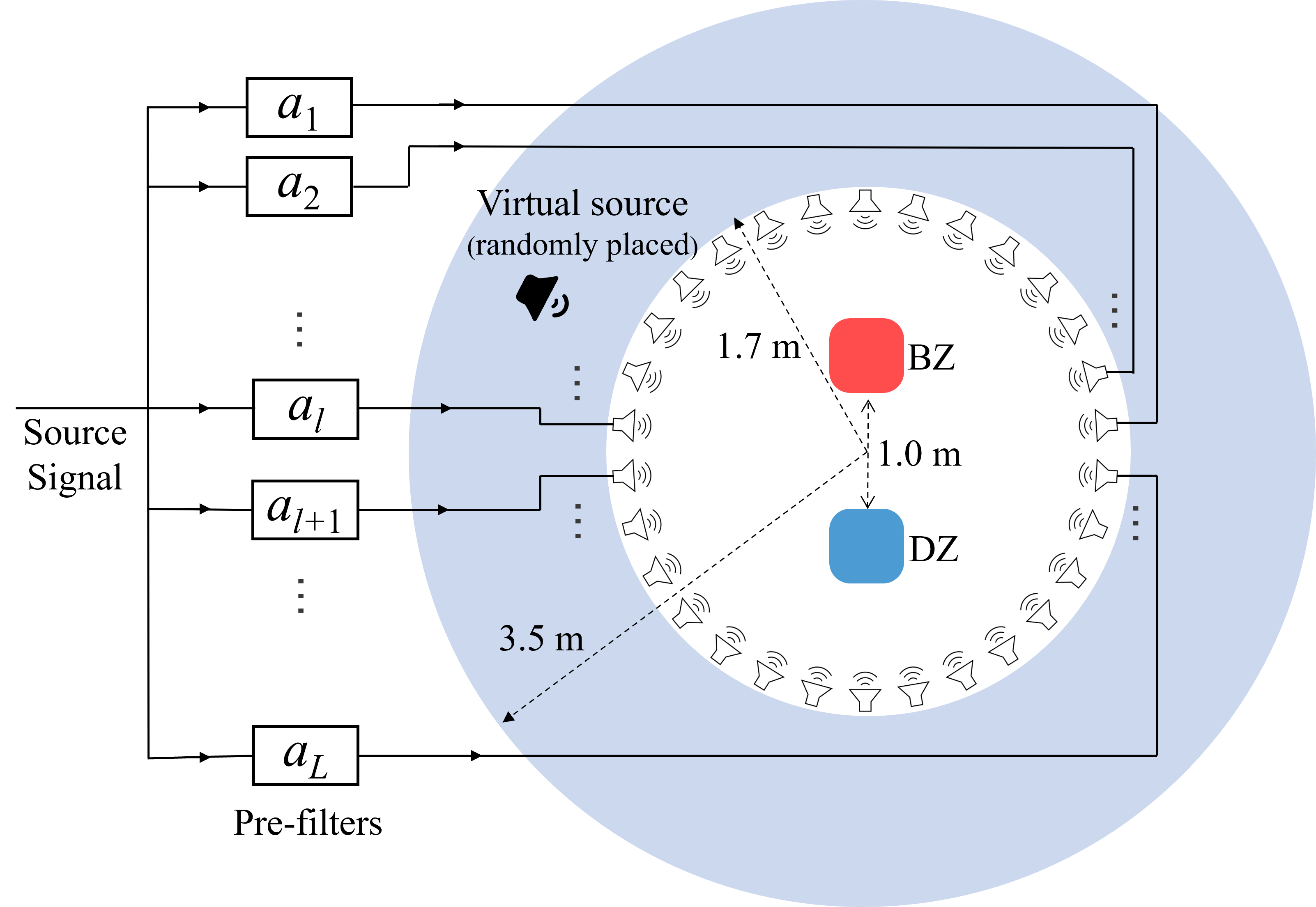}
    \caption{Configuration of the PSZ system.}
    \label{fig:PSZconfig}
\end{figure}

% \begin{figure}[htb]
% \label{fig:PSZconfig}
% \begin{minipage}[b]{1.0\linewidth}
%   \centering
%   \centerline{\includegraphics[width=8cm]{PSZsystem.png}}
% %  \vspace{2.0cm}
%   \centerline{(a) Personal sound zone system}\medskip
% \end{minipage}
% \hfill
% \begin{minipage}[b]{1.0\linewidth}
%   \centering
%   \centerline{\includegraphics[width=4cm]{sound zone.jpg}}
% %  \vspace{1.5cm}
%   \centerline{(b) sound zone}\medskip
% \end{minipage}

% %
% \caption{Configuration of a PSZ system.}
% \label{fig:res}
% %
% \end{figure}

Fig.~\ref{fig:PSZconfig} shows a PSZ system formed by an array composed of $L$ loudspeakers. A bright zone and a dark zone are separately defined, normally each with a size being slightly larger than the human head. We use $M_{B}$ and $ M_{D}$ to denote the number of control points respectively set inside the bright zone and the dark zone, and set $M=M_{B}+M_{D}$. The emitting sound of loudspeakers is applied with pre-filters $a_{l}(n), l=1,\dots,L$ to control the sound pressure received at control points $y^{m}, m=1,\dots,M$, where $n, l, m$ are the indices of time sample, loudspeaker and control point, respectively. 

Let $h^{l, m}(n)$ denote the RIR from loudspeaker $l$ to control point $m$, and $s(n)$ denote the source sound to be emitted with loudspeakers. The reproduced sound pressure at the $i$-th control point can be represented in the form of time-domain convolution as $y^{m}(n) = (\sum_{l=1}^{L} h^{l, m}(n) \ast a^{l}(n)) \ast s(n)$. Since the PSZ system can be considered as a linear time invariant (LTI) system, the reproduced acoustic transfer function (ATF) received at the $m$-th control point from the pre-filtered loudspeaker array can be expressed as ${g}^{m}(n) = \sum_{l=1}^{L} a^{l}(n)*h^{m, l}(n)$, which can be transformed to frequency domain as $G^{m}(k) =  \sum_{l=1}^{L} H^{m, l}(k) A^{l}(k)$.
% \begin{equation}
% \label{eq:PMfreq}
%     G^{m}(k) =  \sum_{l=1}^{L} H^{m, l}(k) A^{l}(k).
% \end{equation}
Since the reproduction is conducted in the frequency domain the remainder of this paper, the frequency $k$ will no longer be noted. And the received ATF can be expressed in matrix form as ${\bf g} =  {\bf H}{\bf a}$, 
% \begin{equation}
%     {\bf g} =  {\bf H}{\bf a},
% \end{equation}
where ${\mathbf g}$ is an $M\times1$ vector denoting reproduced ATFs at the microphone array, and ${\bf a}$ is an $L\times1$ vector denoting pre-filters for the loudspeaker array. ${\mathbf H}$ is an $M\times L$ matrix whose $(m, l)$-th element equals to $H^{m, l}$. 
% PM
% The goal of PSZ system pre-filter designing is to reproduce given target ATF $\tilde{\bf g}_B \in C^{M_B\times1}$ for control points in the BZ while constrain the acoustic energy in the dark zone to a low level, which also means the target ATF in DZ can be denotes as a zero vector ${\bf 0} \in C^{M_D\times 1}$. Hence the desired ATFs can be denotes as $\tilde{\bf g} = [\tilde{\bf g}_B, {\bf 0}]$. Therefore, the optimization problem of PM method can be interpreted as: minimizing the error between the desired ATF $\tilde{\mathbf{g}}$ and reproduced ATF ${\bf g}$, namely
% by applying proper pre-filters, namely:
% \begin{equation}
% 	\label{PM-CTF_opti}
% 	\min \limits_{\mathbf{a}} \| \mathbf{Ha} - \tilde{\mathbf{g}} \| ^{2}
%     + \delta \| \mathbf{a} \|^{2}.
% \end{equation}	
% The Tikhonov regularization term with regularization factor $\delta$ is added to avoid ill-conditioned problems in matrix inversion, while also controlling the energy of the pre-filter to ensure the stability of the PSZ system output. The reproduced pre-filter can be obtained as:
% \begin{equation}
%     \label{eq:PMsolve}
% 	\mathbf{a} = (\mathbf{H}^{H} \mathbf{H} + 
%     \delta \mathbf{I})^{-1} \mathbf{H}^{H} \tilde{\mathbf{g}},
% \end{equation}	
% where $H$ denotes the Hermitian transpose.

\section{Neural PSZ method}

\subsection{Neural PSZ system pipeline}
% Fig: DNN pipeline
\begin{figure}[t]
    \centering
    \includegraphics[width=.9\linewidth]{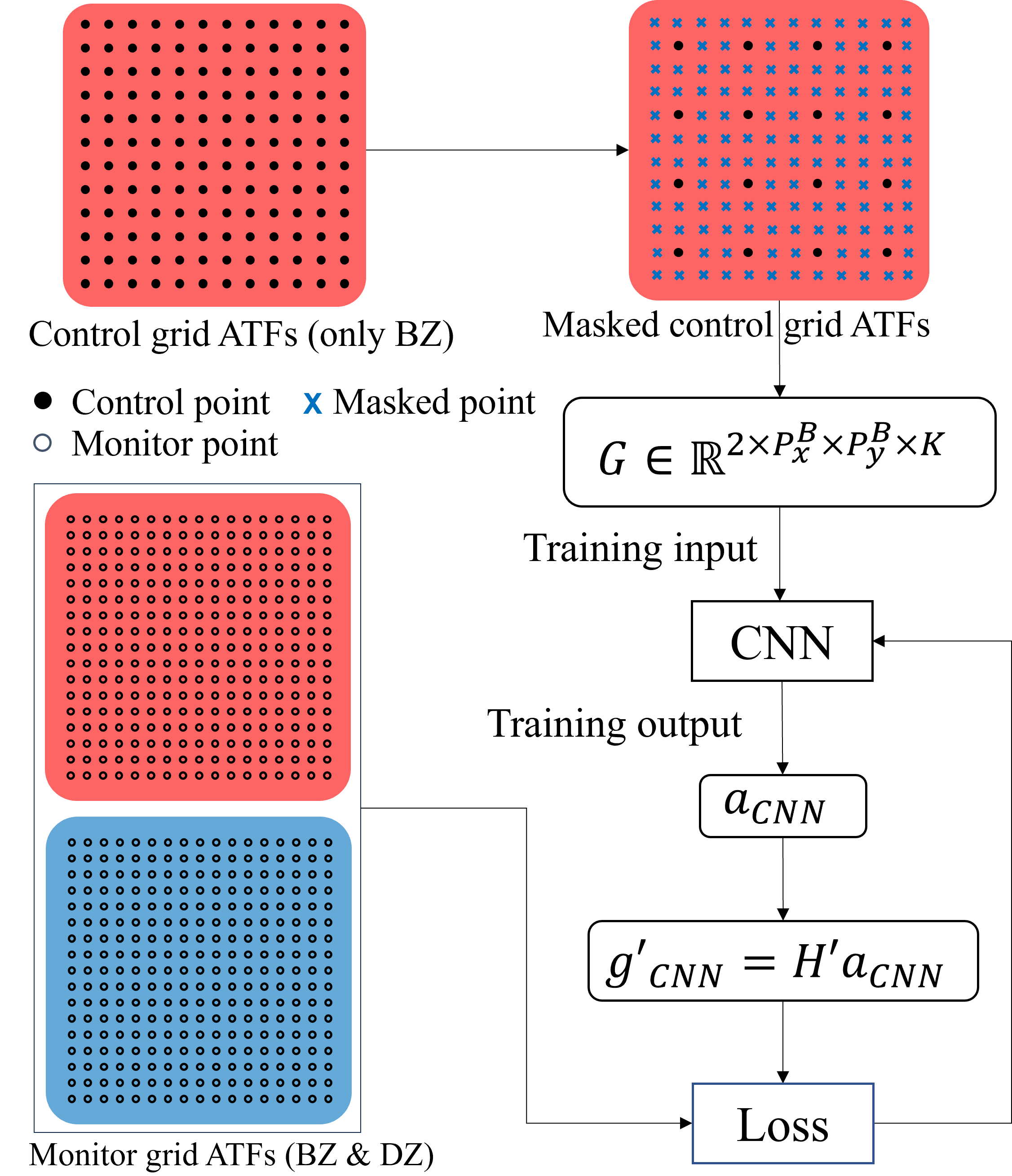}
    \vspace{-10pt}    
    \caption{Pipeline of the proposed Neural PSZ. The target ATFs $\tilde{\mathbf{g}}$ on BZ control grid is masked to a randomly selected grid pattern and then put into the CNN network for training. The output pre-filter set is used to generate the reproduced ATFs ${\mathbf g'}$ on monitor point grid and then compared with ground truth $\tilde{\mathbf g}'$. The masked pattern exemplified here is a $4\times4$ grid with an interval of 3 control points.}
    \label{fig:pipeline}
    
\end{figure}

% Fig: NN arch
%\begin{figure*}[htb]
%\begin{minipage}[b]{.8\linewidth}
%  \centering
%  \centerline{\includegraphics[width=12cm]{network.png}}
% % \vspace{-5pt}
%  \centerline{(a) The proposed CNN architecture.}\medskip
%\end{minipage}
%\hfill
%\begin{minipage}[b]{.18\linewidth}
%  \centering
%  \centerline{\includegraphics[width=4.5cm]{basic block.png}}
%% \vspace{-5pt}
%  \centerline{(b) Residual Block.}\medskip
%\end{minipage}
\begin{figure*}[t]
    \centering
    \includegraphics[width=.98\linewidth]{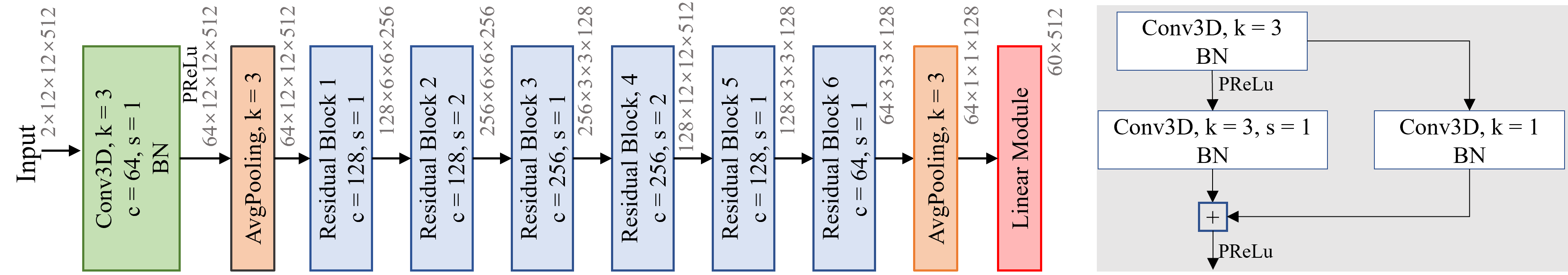}
    \vspace{-10pt}
    %\caption{Configuration of the proposed Neural PSZ network (left) and the inner structure of the residual block (right). The size of input data is consistent with the setting in Sec.~\ref{sec:experiment}}
    \caption{Configurations of the proposed Neural PSZ network (left) and the residual block (right). The size of input is consistent with Sec.~\ref{sec:experiment}.}
    \vspace{-10pt}
    \label{fig:NN}
\end{figure*}

% target
In our proposed end-to-end Neural PSZ approach, the neural network is used to map the desired ATF in PSZs to the pre-filter set for the loudspeaker array. Based on the concept of target matching which is similar to pressure matching, our objective is to minimize the reproduction error between the reproduced ATFs ${\mathbf g}$ and the desired target ATFs $\tilde{\mathbf g}$. The sound scene is only needed to be reproduced in BZ denoted as $\tilde{\bf g}_B \in C^{M_B\times1}$, while the acoustic energy should be constrained to a low level in DZ, hence the target ATF in DZ can be denoted as a zero vector ${\bf 0} \in C^{M_D\times 1}$. Therefore, the target ATFs of the PSZ system can be denoted as $\tilde{\bf g} = [\tilde{\bf g}_B, {\bf 0}]$. 
% monitor grid
To ensure that the neural network consistently captures global spatial information in the zones instead of overfitting to the control points, we introduce a monitor point grid array in each sound zone that does not overlap with the control point grid. The target ATFs $\tilde{\mathbf{g}}' \in \mathbb{C}^{M' \times 1}$ at these monitor points are not provided as inputs to the network ($M'$ denotes the number of monitor points). Instead, the RIRs from the loudspeaker array to the monitor point grid ${\bf H'} \in {\mathbb{C}}^{M'\times L}$ are used with $\tilde{\mathbf{g}}'$ to compute the MSE loss after the network generates the pre-filters. If we define the pre-filter set with $K$ frequency components generated by CNN as $\mathbf{a}_{CNN}(k), k=1, ..., K$, the loss function can be formulated as:
\begin{equation}
    \mathcal{L} = \frac{1}{M'\times K} \sum_{k=1}^{K}
        \Vert \mathbf{H'}(k)\mathbf{a}_{CNN}(k) 
        - \tilde{\mathbf{g}}'(k) \Vert_2
\end{equation}

% Mask
 Meanwhile, to ensure that the network has the ability to learn spatial informations from  different control microphone grid patterns, also to extract the spatial cues from as few control points as possible, we preprocess the input data using different types of grid masks. The ATF data at masked grid points are multiplied by zero, while the unmasked data remain unchanged, therefore the pattern of control point grid fed to the network can be varied. After this procedure, the masked data are then fed into the neural network and directly outputs the pre-filters set for loudspeaker array. Finally the generated pre-filters set is used to compute the actual sound field at the monitoring points and evaluate the loss against the target sound field. The pipeline of the proposed network is shown in Fig.~\ref{fig:pipeline}.

% data representation 
% \vspace{-1em}
% \subsection{Data representation}
For the purpose of facilitating the neural network training, we assume that the control point grid is uniform, namely microphones are evenly spaced in the zone. Under this condition, the control microphone array in each PSZ can be described as a rectangular grid of length $P_{x}$ and height $P_{y}$, and $M_{B, D} = P_{x}^{B, D}\times P_{y}^{B, D}$. Since the function of DZ is always to remain dark/silent regardless of changes in the desired ATF of BZ, we input only the target ATF of BZ $\tilde{\mathbf{g}_{B}}$ into the neural network to avoid data redundancy. Meanwhile, we want to preserve the spatial distribution of the data as completely as possible, therefore the desired frequency response $\tilde{\mathbf{g}}_{B} \in \mathbb{C}^{M_B \times 1}$ for each frequency component $k$ is firstly transformed to a matrix $\hat{\bf G}_{k} \in \mathbb{C}^{P_{x}^{B} \times P_{y}^{B}}$ which is consistent with microphone grid distribution. Considering the discrete set of $K$ frequencies, $\hat{\bf G}_{k}$ is then stacked along an additional dimension to get the full band desired ATF tensor of control point grid, $\hat{\bf G} \in \mathbb{C}^{P_{x}^{B}\times P_{y}^{B}\times{K}}$. In order to train the networks in real domain, we need to transform the complex-valued data to real-valued data, hence we add an extra dimension of length 2 to store its real and imaginary part respectively. Finally, we can get the neural network input as ${\bf G} \in \mathbb{R}^{2\times P_{x}^{B}\times P_{y}^{B}\times K}$, as the illustration in Fig.~\ref{fig:pipeline}. Similarly, we can define the output of the network as tensor ${\bf A} \in \mathbb{R}^{2\times L\times K}$.

% mask
\vspace{-1em}
\subsection{Network architechture}

% Network
We use a 3D convolutional network based on Residual Network (ResNet) architecture \cite{he2015resnet} to construct this end-to-end model. ResNet is composed of stacked residual blocks with shortcut connections, which facilitates more efficient gradient propagation during backpropagation, thereby alleviating the vanishing gradient problem, leading to improved feature representation and performance. 

The architecture of proposed method and the residual block are illustrated in the Fig.~\ref{fig:NN}. We adopt a ResNet-like basic module structure, using PReLU as the activation function, and employ 3D convolutional layers to process the input 4D tensors. In the final linear module, two fully connected layers are sequentially applied to the frequency dimension and channel of output, recovering the output dimension to $\mathbb{R}^{2L\times K}$ ($2L$ denotes $L$ loudspeaker channel with real and image components) and then reshaped to $\mathbf{a}_{CNN}$. Group linear layer is applied to the fully connected layer along channel to ensure that each frequency is assigned its own set of fully connected parameters.

\section{Experiment}
\label{sec:experiment}

\subsection{Experiment setup}
% room
To ensure the authenticity of the data, we conducted the experiments in a reverberant environment with reverberation time (RT60) $250$ ms. The ATF datasets are all simulated using gpuRIR generator \cite{Diaz2020gpuRIR} based on the image source method. As shown in Fig.~\ref{fig:PSZconfig}, the PSZ system for experimental evaluation comprises a circular array of 30 evenly distributed loudspeakers in a rectangular room with dimensions $8\times 8 \times3 m^{3}$. The radius of the circular loudspeaker array is 1.68 m, and the BZ and DZ are both $0.4\times0.4 m^{2}$ approximating the upper limit of human head size, with an interval of $1 m$ between two zones. In each sound zone, the width $P_x^{B, D}$ and height $P_y^{B, D}$ of the control point grid are set to 12, therefore a total of 144 evenly distributed control microphones with 3.64 cm spacing. Similarly, for the monitor point grid,  ${P'}_x^{B, D} = {P'}_y^{B, D} = 17$, hence each zone has 289 monitor points evenly distributed with 2.5 cm spacing. Note that the spatial Nyquist frequencies of the loudspeaker array and control microphone array are 483 and 4250 Hz, respectively. Referring to the reproduce target setup in \cite{hong2023reproduction}, we define the reproduced target as a virtual sound source randomly positioned in the room with its location constrained to lie within an annular region centered at the origin, and its radius range is [1.7, 3.5] m, considering the center of the room as the coordinate origin. 20,000 pairs of ATFs in BZ on both control point grid and monitor point grid with 512 frequency components on the frequency range of $[0, 2000]$ Hz are used as network input and ground-truth respectively.

% Grid mask
In the experiment, 10 types of different masking grid pattern are selected. All of these grid patterns are selected on the control microphone grid with different microphone numbers and different intervals. Utilizing the raw grid with no data masking as the initial grid pattern Grid-12, and its control point spacing as unit interval, the masked grid patterns used for training are listed: 
(1) Grid-12: $12\times 12$ with no mask. 
(2) Grid-6: $6\times 6$ with the interval of 2 unit intervals. 
(3) Grid-4: $4\times 4$ the interval of 3 unit intervals. 
(4) Grid-3\#1: $3\times 3$ the interval of 4 unit intervals. 
(5) Grid-3\#2: $3\times 3$ the interval of 3 unit intervals. 
(6) Grid-3\#3: $3\times 3$ the interval of 2 unit intervals. 
(7) Grid-2\#1: $2\times 2$ the interval of 6 unit intervals. 
(8) Grid-2\#2: $2\times 2$ the interval of 4 unit intervals. 
(9) Grid-2\#3: $2\times 2$ the interval of 1 unit interval. 
(10) Grid-1: single point in the centre. 
% train
For network training, the Adam optimizer is used with a learning rate of 0.001. The training process is usually time-consuming, but it can often be operated offline. The model size is 21.59M parameters, and it is calculated using a NVIDIA V100 TENSOR CORE GPU (graphics processing unit, NVIDIA Corp., Santa Clara, CA).

\vspace{-5pt}
\subsection{ Baseline and Metrics}
\textbf{Baseline method.} 
The PM method \cite{druyvesteyn1997personal} is used as a baseline. Its optimization objective is to minimize the error between reproduced ATFs in PSZs and the target one. The target as input for PM method in this section is the target ATF input of masked control points grid, which is consistent with Neural PSZ method. 

% The optimization problem can be formulated as
% \begin{equation}
% 	\label{PM_opti}
% 	\min \limits_{\mathbf{a}} \| \mathbf{Ha} - \tilde{\mathbf{g}} \| ^{2}
%     + \delta \| \mathbf{a} \|^{2},
% \end{equation}	
% and the Tikhonov regularization term with regularization factor $\delta$ is added to avoid ill-conditioned problems in matrix inversion, while also controlling the energy of the pre-filter to ensure the stability of the PSZ system output. The reproduced pre-filter can be obtained as:
% \begin{equation}
%     \label{eq:PMsolve}
% 	\mathbf{a} = (\mathbf{H}^{H} \mathbf{H} + 
%     \delta \mathbf{I})^{-1} \mathbf{H}^{H} \tilde{\mathbf{g}},
% \end{equation}	
% where $H$ denotes the Hermitian transpose.
\textbf{Metrics.} \textbf{(i) Relative mean energy error (RE)} is defined as the ratio of the error between the reproduced ATF and the desired ATF to the mean energy of the desired ATF target, and is presented in the dB scale. Since the target ATF for RE in DZ is a zero vector, we still use the mean energy of target ATF in BZ as a reference sound pressure to ensure the consistency in the relative of magnitudes of the RE values between two zones. For RE, the smaller the better.
\textbf{(ii) Acoustic contrast (AC)} is the ratio of spatially averaged pressures at the given frequency between the bright zone and the dark zone. The broadband version of AC is denoted as \textbf{bAC}. For AC, the larger the better.
\textbf{(ii) Array Effort (AE)} describes the energy cost of the total loudspeaker array relative to a single reference source loudspeaker rendering the same pressure in the bright zone. The broadband AE is denoted as \textbf{bAE}. For a more detailed understanding of AC and AE metrics, readers can refer to our previous work \cite{tang2025relax}.
All metrics are measured at the monitoring point grid, as our primary focus is on the overall performance within the PSZs.
 
\vspace{-5pt}
\subsection{Results}
% \usepackage{booktabs}
\begin{figure}
    \centering
    \includegraphics[width=\linewidth]{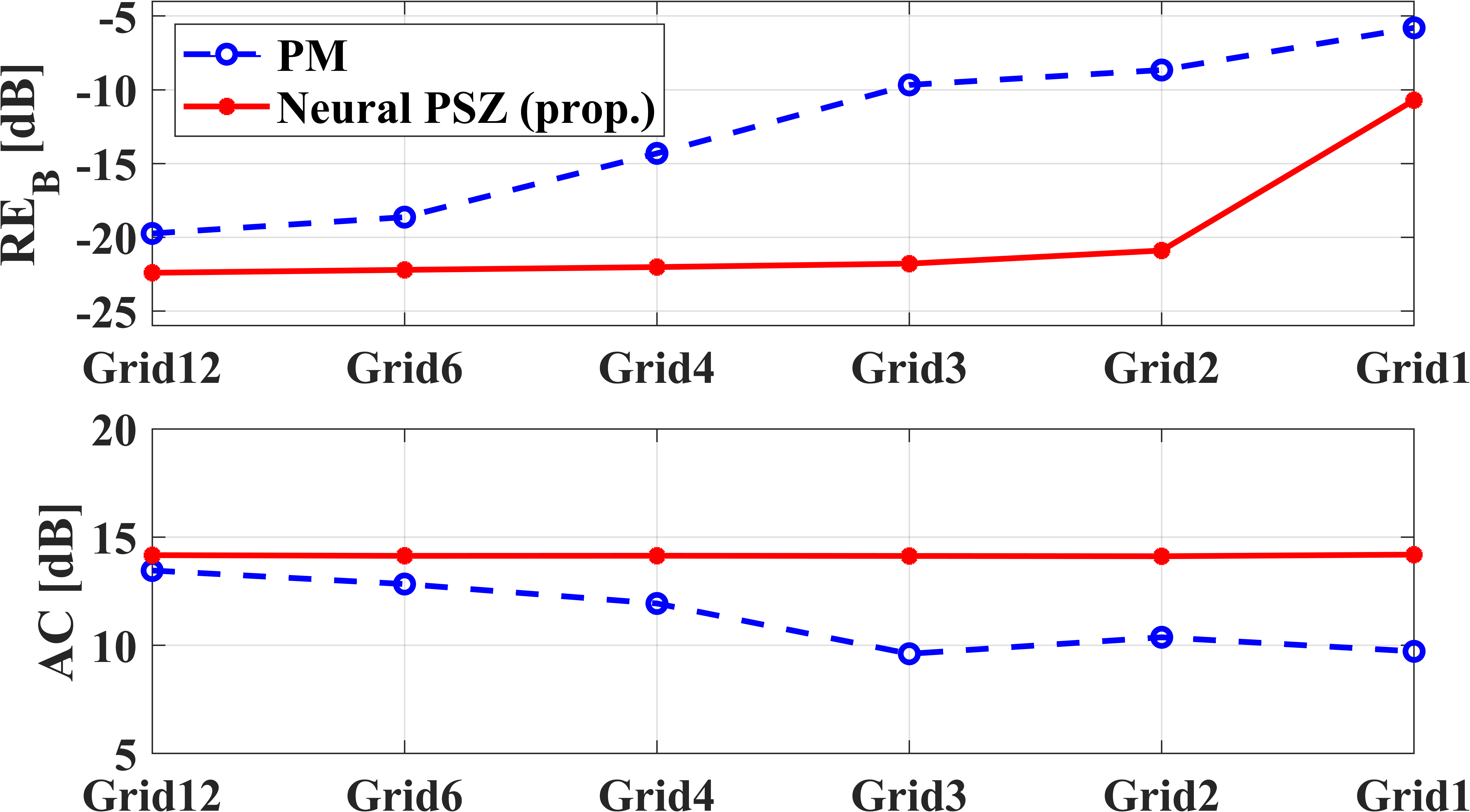}
    \caption{Comparison of $RE_B$ and AC for PM and Neural PSZ.}
    \label{fig:grid metrics}
\end{figure}

% \usepackage{booktabs}
\begin{table}[]
\centering
\resizebox{0.4\textwidth}{!}{  % 宽度占 ？% 文本宽度，高度自动
    \begin{tabular}{@{}cccc|ccc@{}}
    \toprule
              & \multicolumn{3}{c|}{PM}              & \multicolumn{3}{c}{Neural PSZ   (prop.)} \\ \midrule
        & $RE_B$↓ & $RE_D$↓ & AC↑ & $RE_B$↓   & $RE_D$↓  & AC↑  \\
        Grid-3\#1 & -9.67       & -17.25       & 9.61    & -21.79        & -33.36         & 14.12     \\
        Grid-3\#2 & -9.87       & -17.23       & 9.13    & -21.86        & -33.33        & 14.12     \\
        Grid-3\#3 & -8.70       & -16.39       & 7.73    & -21.87        & -33.32         & 14.12     \\ \bottomrule
    \end{tabular}
    }
\caption{Comparisons of $RE_B$, $RE_D$ and AC between PM and Neural PSZ on different $3 \times 3$ control grid patterns.}
\label{tab:grid pattern}
\end{table}

\begin{table}[]
\centering
\resizebox{0.4\textwidth}{!}{  % 宽度占 80% 文本宽度，高度自动
    \begin{tabular}{@{}cccc|ccc@{}}
\toprule
          & \multicolumn{3}{c|}{Flexible grid} & \multicolumn{3}{c}{Fixed grid} \\ \midrule
   & $RE_B$↓      & $RE_D$↓      & AC↑            & $RE_B$↓       & $RE_D$↓  & AC↑   \\
Grid-12   & -22.41     & -32.16      & 14.17 & -22.67      & -32.70       & 14.07 \\
Grid-6    & -22.21     & -32.94      & 14.13 & -22.68      & -32.66       & 14.08 \\
Grid-4    & -22.03     & -33.11      & 14.14 & -22.64      & -32.64       & 14.08 \\
Grid-3\#1 & -21.79     & -33.36      & 14.13 & -22.60      & -32.69       & 14.06 \\
Grid-2\#1 & -20.90     & -33.76       & 14.12 & -22.18      & -33.09       & 14.05 \\ \bottomrule
\end{tabular}
}
\caption{Comparison of $RE_B$, $RE_D$ and AC for Neural PSZ network based on flexible-grid training and fixed-grid training.}
% \vspace{-5pt}
\label{tab:flexible&fix}
\end{table}

% Fig: Real part
\begin{figure}[htb]
\centering
    \includegraphics[width=\linewidth]{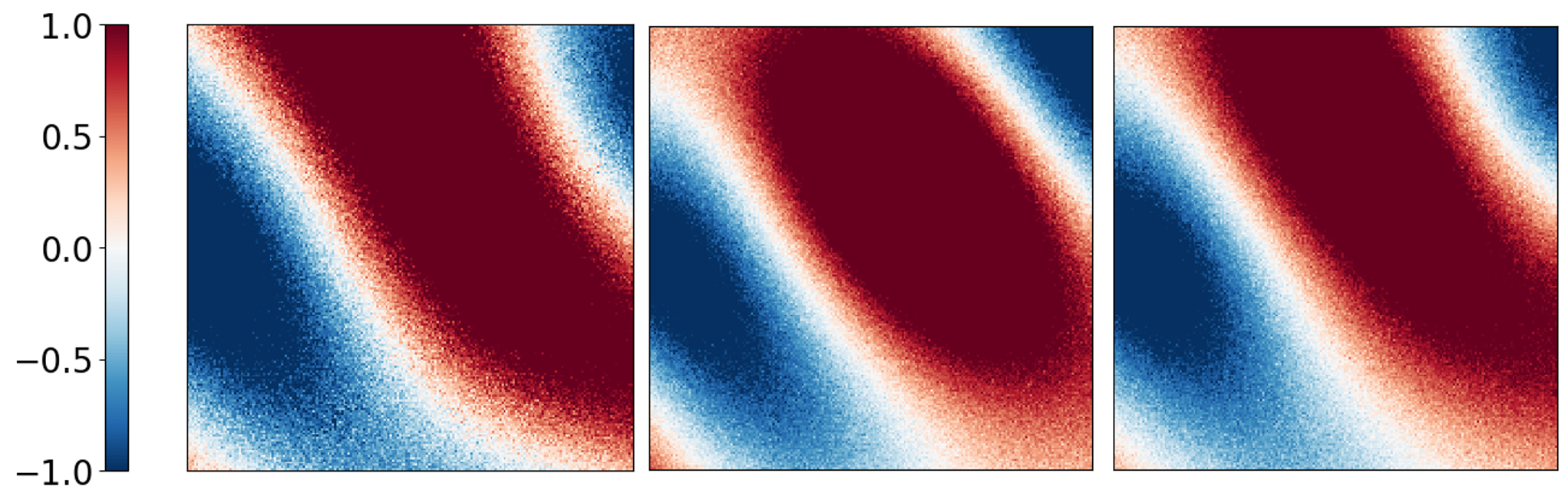}
    % \caption{Configuration of the PSZ system.}
\vspace{-15pt}
%\caption{Comparison of BZ reproduced ATF (real part) distributions with a virtual source at (1.2, 1.8) as target at 875 Hz. From left to right, the figures show the ATF of ground truth (left), the ATF reproduced by PM method (middle), and the ATF reproduced by the proposed Neural PSZ method (right). Both methods use Grid-3\#1. }
\caption{Comparisons of BZ reproduced ATF (real part) with a virtual source located at (1.2, 1.8): (left) ground truth, (middle) the PM method, (right) the proposed Neural PSZ method. Both methods use Grid-3\#1. The frequency is 875 Hz.}
\label{fig:real part}
\end{figure}

% \begin{figure}[htb]
% \centering
% \begin{minipage}[b]{.25\linewidth}
%   \centering
%   \centerline{\includegraphics[height=2.7cm]{real_target_B.png}}
%     \vspace{-0.1cm}
%   \centerline{(a) Target (BZ).}\medskip
% \end{minipage} %
% \hfill
% \hspace{0.02\textwidth}
% \begin{minipage}[b]{.25\linewidth}
%   \centering
%   \centerline{\includegraphics[height=2.5cm]{real_PM_B.png}}
% %  \vspace{2.0cm}
%   \centerline{(b) PM (BZ).}\medskip
% \end{minipage}%
% \hfill
% \hspace{0.02\textwidth}
% \begin{minipage}[b]{.25\linewidth}
%   \centering
%   \centerline{\includegraphics[height=2.5cm]{real_dnn_B.png}}
% %  \vspace{2.0cm}
%   \centerline{(c) Neural PSZ (BZ)}\medskip
% \end{minipage}%
% \vspace{-15pt}
% \caption{Reproduced ATF (real part) distributions with a virtual source at (1.2, 1.8) as target at 875 Hz. Both methods use Grid-3\#1. }
% \label{fig:real part}
% %
% \end{figure}

% Exp 1: baseline comparison
First, we compared the performance of RE in BZ and AC score and its change with varied grid pattern of PM method and our proposed Neural PSZ method. The Tikhonov regularization factor of the PM method is globally searched and tuned to ensure that its AE performance on the monitor point grid matches that of the Neural PSZ method on the test set on average, thereby ensuring a fair comparison.
As Fig.~\ref{fig:grid metrics} shows, the performance of the PM method gradually decreases as the input grid becomes sparser, since the number of points for matching and controlling is gradually decreasing. For the Neural PSZ method, the performance of both metrics is kept better than PM with slight decrease as the grid becomes sparser, while the RE scores sharply increase with Grid-1 pattern, where only a single-point ATF is considered as input. Unlike the multiple-point control point grid which contains relative position information among each point, the single-point ATF contains nearly no spatial information. This indicates that the Neural PSZ method can effectively learn the spatial information of the target sound zone from a small number of points, and can use this information to generate pre-filters for PSZ reconstruction.

% Exp 2: random grid
As shown in Table.~\ref{tab:grid pattern}, we also compared the differences of performance between the PM method and the Neural PSZ method under the same number of control points $3\times3$ as the interval of grid spacing varied. With the grid spacing decreasing, which means the control point grid 'contracts' toward the center and its actual coverage area gradually shrinks, the RE and AC performance of the PM method in PSZ region gradually declines, indicating a loss of control in PSZs. In contrast, the performance of the Neural PSZ method remains largely unchanged, indicating that the neural network can still learn and infer global spatial information from these local, limited points. 
We also present the real part of reproduced ATFs in both BZ and DZ, and that of the target ATFs in BZ at 875 Hz for reference, as shown in Figure.~\ref{fig:real part} . Utilizing Grid-3\#1 pattern with an interval of 7.3 cm as input, the PM method can hardly maintain the reproduction of the target sound field in the BZ, especially on the edge of zone. In contrast, the Neural PSZ method is still able to reproduce the BZ sound field relatively completely while keeping the energy in the DZ at a low level.

Finally, we also conducted individual training on some fixed grid patterns, as listed in Table.~\ref{tab:flexible&fix}, and compared the results with the network used in the former paragraphs which is trained on randomly selected grid patterns, or can be called as flexible-grid training. Compared with fixed-grid training, the network based on flexible-grid training incurs some degradation of RE in the BZ. Moreover, as the grid becomes increasingly sparse, this degradation grows more significant, reaching a difference of about 1.3 dB in the case of Grid-2\#1. Nevertheless, we believe this trade-off is acceptable when compared to the advantages of handling variable grid patterns, since the network based on flexible-grid training can adapt to diverse grid configurations without retraining, which means it is more suitable for real-world applications where fixed grids are not guaranteed.

\section{Conclusion}
\label{sec:majhead}
In this paper, a Neural PSZ system based on 3D CNN network for pre-filter designing has been proposed and examined. The proposed method ensures that the system can flexibly utilizing ATFs from varied control microphone grid patterns to reproduce PSZs, and can freely choose the virtual acoustic scene to be reproduced after one single training session, which is highly beneficial for real-world applications. At the same time, the proposed method is able to use less control microphones to reproduce a target acoustic scene in BZ with low RE. Meanwhile, it maintains good performance of AC score between BZ and DZ, which denotes the ability to enhance the degree of separation between each zone. We believe that these performances and characteristics have significant importance and value for the lightweight and flexible application of PSZ system in the future, and the system could adapt to a wider variety of scenarios with greater flexibility in future developments.

% Below is an example of how to insert images. Delete the ``\vspace'' line,
% uncomment the preceding line ``\centerline...'' and replace ``imageX.ps''
% with a suitable PostScript file name.
% -------------------------------------------------------------------------
% \begin{figure}[htb]
% \begin{minipage}[b]{1.0\linewidth}
%   \centering
%   \centerline{\includegraphics[width=8.5cm]{image1}}
% %  \vspace{2.0cm}
%   \centerline{(a) Result 1}\medskip
% \end{minipage}
% %
% \begin{minipage}[b]{.48\linewidth}
%   \centering
%   \centerline{\includegraphics[width=4.0cm]{image3}}
% %  \vspace{1.5cm}
%   \centerline{(b) Results 3}\medskip
% \end{minipage}
% \hfill
% \begin{minipage}[b]{0.48\linewidth}
%   \centering
%   \centerline{\includegraphics[width=4.0cm]{image4}}
% %  \vspace{1.5cm}
%   \centerline{(c) Result 4}\medskip
% \end{minipage}
% %
% \caption{Example of placing a figure with experimental results.}
% \label{fig:res}
% %
% \end{figure}

%\vfill
%\pagebreak

\vfill\pagebreak

\bibliographystyle{IEEEbib}
\bibliography{strings,refs}

\end{document}